\documentclass[12pt]{article}

\usepackage{amssymb,amsmath,amsthm,amscd,latexsym}
\usepackage{amsfonts}
\usepackage{tikz}
\usetikzlibrary{arrows,positioning,shapes,fit,calc}
\usepackage{tikz-cd}
\usepackage[mathscr]{eucal}
\usepackage{color,hyperref}
\usepackage{graphicx}
\usepackage{multirow}
\usepackage[ruled,vlined]{algorithm2e}
 \newtheorem{theorem}{Theorem}[section]
 
 \newtheorem{lemma}[theorem]{Lemma}
 
 \theoremstyle{definition}
 \newtheorem{definition}[theorem]{Definition}
 \theoremstyle{remark}
 
 \newtheorem{example}[theorem]{Example}
 \numberwithin{equation}{section}
\begin{document}

\title{Reversible $G^k$-Codes with Applications to DNA Codes}
\author{ Adrian Korban \\
Department of Mathematical and Physical Sciences \\
University of Chester\\
Thornton Science Park, Pool Ln, Chester CH2 4NU, England \\
Serap \c{S}ahinkaya \\
Tarsus University, Faculty of Engineering \\ Department of Natural and Mathematical Sciences \\
Mersin, Turkey \\
Deniz Ustun \\
Tarsus University, Faculty of Engineering \\ Department of Computer Engineering \\
Mersin, Turkey}
 \maketitle

\begin{abstract}
In this paper, we give a matrix construction method for designing DNA codes that come from group matrix rings. We show that with our construction one can obtain reversible $G^k$-codes of length $kn,$ where $k, n \in \mathbb{N},$ over the finite commutative Frobenius ring $R.$ We employ our construction method to obtain many DNA codes over $\mathbb{F}_4$ that satisfy the Hamming distance, reverse, reverse-complement and the fixed GC-content constraints. Moreover, we improve many lower bounds on the sizes of some known DNA codes and we also give new lower bounds on the sizes of some DNA codes of lengths $48, 56, 60, 64$ and $72$ for some fixed values of the Hamming distance $d.$
\end{abstract}


\section{Introduction}

The interest in studying and designing DNA codes originated in 1994 when L. M. Adleman solved a computationally difficult mathematical problem by introducing an algorithm using DNA strands and molecular biology tools \cite{Adleman}. Since then, many other applications of DNA codes were discovered, such as using synthetic DNA for digital media storage \cite{Blawat1, Church1} or using DNA codes to break a cryptosystem known as DES \cite{Adleman2, Boneh1}. In order to use DNA codes in data storage or in cryptosystems, one needs to consider DNA codes that satisfy certain properties. It is not an easy task to find an effective method for constructing DNA codes that satisfy these conditions and this has attracted interest from many researchers \cite{King1, Limbachiya1, Marathe1}.

Some known methods for designing DNA codes that satisfy certain conditions include the study of reversible self-dual codes over $GF(4)$ \cite{Kim1}, the study of cyclic and extended cyclic constructions \cite{Aboluion1} or the study of linear constructions \cite{GaboritI}. Recently in \cite{ReversibleDNACodes}, linear codes derived from group ring elements are considered to construct reversible DNA codes that satisfy the Hamming distance, reverse, reverse-complement and the fixed GC-content constraints. Moreover, in \cite{ReversibleDNACodes}, many new lower bounds on the sizes of some DNA codes that satisfy the above constraints are found. This suggests that the study of group rings is an interesting research direction that may have some useful applications to DNA coding.

In this work, we look at group matrix rings and employ a matrix construction given in \cite{M(R)GPaper} to show that one can construct reversible codes of length $kn$ with, where $k$ and $n$ are both positive even integers, over the finite commutative Frobenius ring $R.$ We use our matrix construction to obtain many DNA codes over the finite field $\mathbb{F}_4.$ Our DNA codes satisfy the Hamming distance, reverse, reverse-complement and the fixed GC-content constraints. Moreover, we improve many known (in the literature) lower bounds on the sizes of some known DNA codes and we also give new lower bounds on the sizes of some DNA codes of lengths greater than $42.$ In particular, we give new lower bounds on the sizes of DNA codes of lengths $48, 56, 60, 64$ and $72$ for some fixed Hamming distance $d$ and some fixed GC-content constraint.

The paper is organised as follows. In Section~2, we give the basic definitions and results on linear codes, DNA codes, special matrices, group rings, group codes,reversible group codes, group matrix rings and $G^k$-codes. In Section~3, we give our main code construction using reversible $G^k$-codes and we make a connection with DNA codes. Moreover, in Section~3, we present one possible generator matrix using our main code construction method and we employ it to construct many DNA codes with. The DNA codes that we construct satisfy the Hamming distance, reverse, reverse-complement and the fixed GC-content constraints. Some of these DNA codes are optimal and many have better parameters than some known (in the literature) DNA codes. In Section~4, we tabulate our results. We finish with concluding remarks and directions for possible future research.

\section{Preliminaries}

\subsection{Linear Codes and DNA Codes}\label{LinearandDNACodes}

In this section, we recall basic definitions on linear codes and DNA codes.

Let $\mathbf{x}=(x_1,x_2,\dots,x_n),$ where $x_i \in S_{D_4}=\{A,C,G,T\}$ (representing the four nucleotides in DNA, adenine (A), cytosine (C), guanine (G) and thymine (T)). We use a hat to denote the Watson-Crick complement of a nucleotide, $\hat{A}=T, \hat{T}=A, \hat{C}=G$ and $\hat{G}=C.$ A linear code of length $n$ over $\mathbb{F}_4$ is a subspace of $\mathbb{F}_4^n,$ and we also call an element of a linear code a codeword. The Hamming distance $d(\mathbf{x},\mathbf{y})$ between two codewords is the number of coordinates in which $\mathbf{x}$ and $\mathbf{y}$ are distinct. The minimum Hamming distance $d$ of a linear code $C$ is defined as min$\{d(\mathbf{x},\mathbf{y}) \ | \ \mathbf{x} \neq \mathbf{y}, \ \forall \ \mathbf{x}, \mathbf{y} \in C\}.$ Codes with the largest known $d$ for a specific length $n$ are said to be optimal. A DNA code $D$ of length $n$ is defined as a set of codewords $(x_1,x_2,\dots,x_n)$ where $x_i \in S_{D_4}=\{A,T,C,G\},$ such that $D$ satisfies some or all of the following constraints:
\begin{enumerate}
\item[(i)] The Hamming distance constraint (HD): 
$$d(\mathbf{x},\mathbf{y}) \geq d, \ \ \forall \ \mathbf{x}, \mathbf{y} \in D,$$
for some Hamming distance $d.$
\item[(ii)] The reverse constraint (RV):
$$d(\mathbf{x}^r,\mathbf{y}) \geq d, \ \ \forall \ \mathbf{x}, \mathbf{y} \in D, \ \text{including} \ \mathbf{x}=\mathbf{y}$$
for some Hamming distance $d.$
\item[(iii)] The reverse-complement constraint (RC):
$$d(\mathbf{x}^{rc},\mathbf{y}) \geq d, \ \ \forall \ \mathbf{x}, \mathbf{y} \in D, \ \text{including} \ \mathbf{x}=\mathbf{y}$$
for some Hamming distance $d.$
\item[(iv)] The fixed $GC$-content constraint (GC):
The set of codewords with length $n,$ distance $d$ and $GC$ weight $w,$ where $w$ is the total number of $Gs$ and $Cs$ present in the DNA strand:
$$w_{\mathbf{x}_{DNA}}=|\{x_i: \mathbf{x}_{DNA}=(x_i), x_i \in \{C,G\}\}|,$$
\end{enumerate}
where $\mathbf{x}^r=(x_n,x_{n-1},\dots, x_2,x_1)$ is the reverse of a DNA codeword, $\mathbf{x}^c=(x_1^c,x_2^c,\dots, x_n^c)$ is the complement of a DNA codeword and $\mathbf{x}^{rc}=(x_n^c,x_{n-1}^c,\dots, x_2^c,x_1^c)$ is the reverse complement of a DNA codeword. In this paper, the fixed $GC$-content is simply half the length of the DNA code $D.$
 
A DNA code can be identified with a code over $\mathbb{F}_4=\{0,1,\omega,\omega^2\}$ by employing the standard bijective correspondence between $\mathbb{F}_4$ and the DNA alphabet $S_{D_4}=\{A,T,C,G\}$ given by
$$ \eta:  \mathbb{F}_4 \rightarrow S_{D_{4}},$$
with $ \eta(0)=A, $  $ \eta(1)=T,  $ $ \eta(\omega)=C $  and $ \eta(\omega ^{2})=G $. The same correspondence has already been used in the literature, for example, please see \cite{Kim1}. We extend the bijection $\eta$ so that $\eta (C)$ is regarded as a DNA code for some code $C$ over $\mathbb{F}_4.$ 

We denote the complete weight enumerator of a code $C$ over $\mathbb{F}_4$ by
$$CWE_C(a,b,c,d)=\sum_{c \in C} a^{n_0(c)}b^{n_1(c)}c^{n_{\omega}(c)}d^{n_{\omega^2}(c)},$$
where $n_s(c)$ denotes the number of occurrences of $s$ in a codeword $c.$ We identify the complete weight enumerator of a DNA code $D$ with that of a code $C$ over $\mathbb{F}_4,$ where $D=\eta (C).$ The GC-weight of a codeword $c \in C$ is the sum of $n_{\omega}(c)$ and $n_{\omega^2}(c).$ Therefore, if we let
$$GCW_C(a,b)=CWE_C(a,a,b,b),$$
then $GCW_C(a,b)$ is the GC-weight enumerator of a code $C,$ where the coefficient of $b^i$ is the same as the number of codewords with GC-weight $i.$

Let $A_4^R(n,d)$ denote the maximum cardinality of a DNA code for a given distance $d$ and length $n$ that satisfies the Hamming distance and reverse constraints. Let $A_4^{RC}(n,d)$ be the maximum size of a DNA code of length $n$ satisfying the HD and RC constraints for a given d, $A_4^{GC}(n,d,w)$ be the maximum size of a DNA code of length $n$ satisfying the HD constraint for a given $d$ with a constant GC-weight $w,$ and $A_4^{RC,GC}(n,d,w)$ the maximum size of a DNA code of length $n$ satisfying the HD and RC constraints for a given $d$ with a constant GC-weight $w.$ In \cite{Marathe1}, for an even $n$, the following equality is given;
\begin{equation}\label{even}
A_4^{RC}(n,d)=A_4^{R}(n,d).
\end{equation}

\subsection{Special Matrices, Group Rings and Group Codes}\label{GroupRingsandGroupCodes}

In this section, we recall the definitions of some special matrices that we use later in this work. We also give the basic definitions of group rings.

A circulant matrix is one where each row is shifted one element to the right relative to the preceding row. We label the circulant matrix as $circ(\alpha_1,\alpha_2,\dots , \alpha_n),$ where $\alpha_i$ are the ring elements appearing in the first row. The transpose of a matrix $A,$ denoted by $A^T,$ is a matrix whose rows are the columns of $A,$ i.e., $(A^T)_{ij}=A_{ji}.$

We shall now give the standard definition of group rings.
Let $G$ be a finite group of order $n$ and let $R$ be a finite ring. Then any element in $RG$ is of the form $v=\sum_{i=1}^n \alpha_i g_i$, $\alpha_i \in R$, $g_i \in G.$
Addition in $RG$ is done by coordinate addition, namely $$\sum_{i=1}^n \alpha_i g_i + \sum_{i=1}^n \beta_i g_i =
\sum_{i=1}^n (\alpha_i + \beta_i ) g_i.$$ The product of two elements in  $RG$ is given by
$$\left(\sum_{i=1}^n \alpha_i g_i\right)\left( \sum_{j=1}^n \beta_j g_j\right)  = \sum_{i,j} \alpha_i \beta_j g_i g_j .$$ This gives that the coefficient of $g_k$ in the product is $ \sum_{g_i g_j = g_k } \alpha_i \beta_j .$

The following matrix construction was given by Hurley in \cite{HurleyI}. The same matrix construction was used to study  group codes over Frobenius rings in \cite{Dougherty1}. Let $%
R$ be a finite commutative Frobenius ring and let $G=\{g_1,g_2,\dots,g_n\}$
be a group of order $n$ and let $v=\sum_{i=1}^n \alpha_{g_i} \in RG.$ Define
the matrix $\sigma(v) \in M_n(R)$ to be 

\begin{equation} \label{equation:construction}
\sigma(v) =
\left(
\begin{array}{ccccc}
\alpha_{g_1^{-1} g_1} & \alpha_{g_1^{-1} g_2} & \alpha_{g_1^{-1} g_3} & \dots & \alpha_{g_1^{-1} g_n}  \\
\alpha_{g_2^{-1} g_1} & \alpha_{g_2^{-1} g_2} & \alpha_{g_2^{-1} g_3} & \dots & \alpha_{g_2^{-1} g_n}  \\
\vdots  & \vdots & \vdots & \vdots & \vdots \\
\alpha_{g_n^{-1} g_1} & \alpha_{g_n^{-1} g_2} & \alpha_{g_n^{-1} g_3} & \dots & \alpha_{g_n^{-1} g_n}  \end{array}
\right).
\end{equation}

We note that the elements $g_1^{-1}, \dots, g_{n}^{-1}$ are simply the elements of the group $G$ given in some order.  This particular order is used because it aids in certain proofs and computations. In \cite{Dougherty1}, the following code construction is given:
\begin{equation}\label{groupcodeconstruction}
\mathcal{C}(v)=\langle \sigma(v) \rangle.
\end{equation}
The code is formed by taking the row space of $\sigma(v)$ over the ring $R.$ Such codes are refereed to as group codes or, for simplicity, $G$-codes. Moreover, in \cite{Dougherty1}, it is shown that this matrix construction of $G$-codes corresponds to an ideal in the group ring $RG$ and thus the resulting group code has the group $G$ as a subgroup of its automorphism group. Please see \cite{Dougherty1} for more details on group codes generated from group rings. From now on, every time we refer to $G$-codes, we mean codes constructed as given above.

\subsection{Reversible Group Codes}

Here, we recall an interesting result from \cite{ReversibleDNACodes} on group codes. Namely, this result shows that for certain groups and for a specific ordering of the group elements, one can construct $G$-codes that are reversible. We first start with a definition from \cite{ReversibleDNACodes}.

\begin{definition}
A code $C$ is said to be reversible of index $\alpha$ if $\mathbf{a}_i$ is a vector of length $\alpha$ and $\mathbf{c}^{\alpha}=(\mathbf{a}_0,\mathbf{a}_1,\dots,\mathbf{a}_{s-1}) \in C$ implies that $(\mathbf{c}^{\alpha})^r=(\mathbf{a}_{s-1},\mathbf{a}_{s-2},\dots,\mathbf{a}_{1},\mathbf{a}_{0}) \in C.$
\end{definition}

Let $ G $ be a finite group of order $n=2l$ and let $H=\{e, h_1, h_2, \dots, h_{\ell-1}\}$ be a subgroup of index $2$ in $G$.
Let $\beta \notin H$ be an element in $G$, with $\beta^{-1}=\beta$. We list the elements of $G=\{g_1, g_2, \dots, g_n\}$ as follows:

\begin{equation}\label{fixed}
\{ e, h_1, \dots, h_{\ell-1}, \beta h_{\ell-1}, \beta h_{\ell-2}, \beta h_2, \beta h_1,\beta\}.
\end{equation}
The following result was proved in \cite{ReversibleDNACodes}.

\begin{theorem}\label{RevGroupCode}
Let $R$ be a finite ring. Let $G$ be a finite group of order $n=2\ell$ and let $H=\{e,h_1,h_2,\dots ,h_{\ell-1}\}$ be a subgroup of index 2 in $G.$ Let $\beta \notin H$ be an element in $G$ with $\beta^{-1}=\beta.$ List the elements of $G$ as in (\ref{fixed}), then any linear $G$-code in $R^n$ (a left ideal in $RG$) is a reversible code of index 1.
\end{theorem}

In \cite{ReversibleDNACodes}, the authors make a connection between reversible $G$-codes and DNA codes, this is because reversibility is a desirable property for DNA codes. In this work, we shall make a similar connection between $G^k$-codes and DNA codes. 

\subsection{Group Matrix Rings and $G^k$-Codes}

In this section, we recall a matrix construction for $G^k$-codes which was first given in \cite{M(R)GPaper}.

Let $R$ be a finite commutative ring and let $G=\{g_1,g_2,\dots,g_n\}$
be a group of order $n$.  We note that  no assumption about the groups commutativity is made.  Let $v=A_{g_1}g_1+A_{g_2}g_2+\dots+A_{g_n}g_n \in M_k(R)G,$ that is, each $A_{g_i}$ is a $k \times k$ matrix with entries from the ring $R.$ Define the block matrix $\sigma_k(v) \in M_n(M_k(R))$ to be

\begin{equation}\label{sigmakv}
\sigma_k(v)=
\begin{pmatrix}
A_{g_1^{-1}g_1} & A_{g_1^{-1}g_2} & A_{g_1^{-1}g_3} & \dots &
A_{g_1^{-1}g_n} \\
A_{g_2^{-1}g_1} & A_{g_2^{-1}g_2} & A_{g_2^{-1}g_3} & \dots &
A_{g_2^{-1}g_n} \\
\vdots & \vdots & \vdots & \vdots & \vdots \\
A_{g_n^{-1}g_1} & A_{g_n^{-1}g_2} & A_{g_n^{-1}g_3} & \dots &
A_{g_n^{-1}g_n}
\end{pmatrix}
.
\end{equation}

We note that the element $v$ is an element of the group matrix ring $M_k(R)G.$ In \cite{M(R)GPaper}, it is shown that the above matrix $\sigma_k(v)$ can be used to make two distinct constructions:  

{\bf Construction 1} 
For a given element $v \in M_k(R)G,$ we define the following code over the matrix ring $M_k(R)$:
\begin{equation}
C_k(v)=\langle \sigma_k(v) \rangle.
\end{equation}
Here the code is generated by taking the all left linear combinations of the rows of the matrix with coefficients in $M_k(R).$ 

{\bf Construction 2} 
For a given element $v \in M_k(R)G,$ we define the following  code over the  ring $R$.
Construct the matrix $\tau_k(v)$ by viewing each element in a $k$ by $k$ matrix as an element in the larger matrix.  
\begin{equation}
B_k(v)=\langle \tau_k(v) \rangle.
\end{equation}
Here the code $B_k(v)$ is formed by taking all linear combinations of the rows of the matrix with coefficients in $R$.  
In this case the ring over which the code is defined is commutative so it is both a left linear and right linear code.

In \cite{M(R)GPaper}, the authors showed that with the second construction, one can obtain interesting codes, ones that could not be obtained from other, more classical constructions. Please see \cite{M(R)GPaper} for more details.

We now give some results about the codes $C_k(v)$ and $B_k(v)$. These results can be found in \cite{M(R)GPaper} and we omit the proofs here.

\begin{lemma}  Let $R$ be a finite Frobenius ring and let $G$ be a group of order $n$.  Let $v \in M_k(R)G$.  
\begin{enumerate}
\item The matrix $\sigma_k(v)$ is an $n$ by $n$ matrix with elements from $M_k(R)$ and the code $C_k(v)$ is a length $n$ code over $M_k(R)$.  
\item The matrix $\tau_k(v)$ is an $nk$ by $nk$ matrix with elements from $R$ and the code $B_k(v)$ is a length $nk$ code over $R$.  
\end{enumerate} 
\end{lemma}  

\begin{theorem}
Define $\Omega : (M_k(R))^n \rightarrow M_k(R)G$ by $\Omega(A_0,A_1,\dots,A_{n-1})=\sum A_ig_i.$ In \cite{M(R)GPaper}, it is shown that $C_k(v)$ is a linear code in $(M_k(R))^n$ if and only if $\Omega(C_k(v))$ is a left ideal $M_k(R)G.$
\end{theorem}

The above theorem tells us that any code $C_k(v)$ has $G$ as a subgroup of its automorphism group, just like $G$-codes described in Section~\ref{GroupRingsandGroupCodes}. There is however a distinction between $G$-codes from Section~\ref{GroupRingsandGroupCodes} and the codes $C_k(v).$ Namely, the codes $C_k(v)$ are generated over the matrix ring $M_k(R)$ while $G$-codes are generated over the finite commutative Frobenius ring $R.$ For this reason, to make this distinction clear, we refer to $C_k(v)$ codes as group matrix ring codes. The following result can also be found in \cite{M(R)GPaper}.

\begin{theorem}
The orthogonal of a group matrix ring code $C_k(v)$ is also a group matrix ring code.
\end{theorem}

We now recall some known results about the codes $B_k(v)$. These again can be found in \cite{M(R)GPaper}.

\begin{definition}
Let $G$ be a finite group of order $n$ and $R$ a finite Frobenius commutative ring.   Let  $D$ be a code in $R^{sn}$ where the coordinates can be partitioned into $n$ sets of size $s$ where each set is assigned an element of $G$.
If the code $D$ is held invariant by the action of multiplying the coordinate set marker by every element of $G$ then the code $D$ is called a quasi-group code of length $ns$ and of index $s$.
\end{definition}

\begin{lemma} \label{firstlemma}  Let $R$ be a finite Frobenius ring and let $G$ be a finite group with $v \in M_k(R)$. Then
$B_k(v)$ is a quasi-$G$-code of length $nk$ and index $k$.
\end{lemma} 

Consider a quasi-$G$-code of index $k$.  Then rearranging the coordinates so that the $i$-th coordinates of each group of $k$ coordinates are placed sequentially, then it is easy to see that any $(g_1,g_2,\dots,g_n) \in G^n$ holds the code invariant.  Namely, any quasi-$G$-code of length $kn$ and index $k$ is a $G^k$-code.  This gives the following.

\begin{theorem}Let $R$ be a finite Frobenius ring and let $G$ be a finite group with $v \in M_k(R)$. Then
$B_k(v)$ is a $G^k$ code of length $kn.$  
\end{theorem}

\section{Reversible Group Matrix Ring Codes and Reversible $G^k$-Codes with Applications to DNA Codes}

In this section, we show that for a specific choice for the $k \times k$ matrices $A_{g_i}$ in the matrix $\sigma_k(v)$ given in Equation~(\ref{sigmakv}), one can construct reversible $G^k$ codes of index 1 in $R^{kn}.$ We then make a connection between these codes and DNA codes.

\subsection{The Block Matrix $\sigma_k(v)$ with fixed $k \times k$ Matrices}

We fix the matrices $A_{g_i}$.

Let $R$ be a finite commutative Frobenius ring and let $G=\{g_1,g_2,\dots,g_n\}$
be a group of order $n=2\ell$. Also, let $T_1, T_2,\dots, T_n$ be finite groups, each of order $k=2m.$  We note that  no assumption about the groups commutativity is made. Let $v_{g_i}=\alpha_{(t_i)_1}(t_i)_1+\alpha_{(t_i)_2}(t_i)_2+\dots+\alpha_{(t_i)_k}(t_i)_k \in RT_i.$ Next, let $v=\sigma(v_{g_1})g_1+\sigma(v_{g_2})g_2+\dots+\sigma(v_{g_n})g_n \in M_k(R)G,$ that is, each $\sigma(v_{g_i})$ is a $k \times k$ matrix with the entries from the ring $R.$ Define the block matrix $\sigma_k^*(v) \in M_n(M_k(R))$ to be

\begin{equation}\label{newsigmakv}
\sigma_k^*(v)=
\begin{pmatrix}
\sigma(v_{g_1^{-1}g_1}) & \sigma(v_{g_1^{-1}g_2}) & \sigma(v_{g_1^{-1}g_3}) & \dots &
\sigma(v_{g_1^{-1}g_n}) \\
\sigma(v_{g_2^{-1}g_1}) & \sigma(v_{g_2^{-1}g_2}) & \sigma(v_{g_2^{-1}g_3}) & \dots &
\sigma(v_{g_2^{-1}g_n}) \\
\vdots & \vdots & \vdots & \vdots & \vdots \\
\sigma(v_{g_n^{-1}g_1}) & \sigma(v_{g_n^{-1}g_2}) & \sigma(v_{g_n^{-1}g_3}) & \dots &
\sigma(v_{g_n^{-1}g_n})
\end{pmatrix}
.
\end{equation}

We  make two distinct constructions.  

{\bf Construction 1} 

For a given element $v \in M_k(R)G,$ we define the following code over the matrix ring $M_k(R)$:
\begin{equation}
C_k^*(v)=\langle \sigma_k^*(v) \rangle.
\end{equation}
Here the code is generated by taking the all left linear combinations of the rows of the matrix with coefficients in $M_k(R).$ 

{\bf Construction 2} 

We now define the following  code over the  ring $R$.
Construct the matrix $\tau_k^*(v)$ by viewing each element in a $k$ by $k$ matrix as an element in the larger matrix.  
\begin{equation}
B_k^*(v)=\langle \tau_k^*(v) \rangle.
\end{equation}
Here the code $B_k^*(v)$ is formed by taking all linear combinations of the rows of the matrix with coefficients in $R$.

We next show that for a particular listings of the elements of the groups $G$ and $T_i,$ for all $i=\{1,2,\dots,n\},$ the code $B_k^*(v)$ is a reversible code of length $kn.$

\subsection{Reversible Group Matrix Ring Codes and Reversible $G^k$-Codes}

In this section, we give our method construction that we later use to construct DNA codes with.

Let $ G $ be a finite group of order $n=2l$ and let $H=\{e, h_1, h_2, \dots, h_{\ell-1}\}$ be a subgroup of index $2$ in $G$.
Let $\beta \notin H$ be an element in $G$, with $\beta^{-1}=\beta$. We list the elements of $G=\{g_1, g_2, \dots, g_n\}$ as in Equation~(\ref{fixed}).

Similarly, Let $ T_i $ be a finite group of order $k=2m$ and let 
$S_i=\{e, (s_i)_1,\\
(s_i)_2, \dots, (s_i)_{m-1}\}$ be a subgroup of index $2$ in $T_i$.
Let $\mu \notin S_i$ be an element in $T_i$, with $\mu^{-1}=\mu$. We list the elements of $T_i=\{(t_i)_1, (t_i)_2, \dots, (t_i)_k\}$ as follows:

\begin{equation}\label{fixed2}
\{ e, (s_i)_1, \dots, (s_i)_{m-1}, \mu (s_i)_{m-1}, \mu (s_i)_{m-2}, \mu (s_i)_2, \mu (s_i)_1,\mu\}.
\end{equation}

\begin{theorem}\label{RevGroupMatrixRingCode}
Let $R$ be a finite commutative Frobenius ring, $k$ a positive integer and $G$ be a finite group of order $n=2\ell.$ Also, let $H=\{e,h_1,h_2,\dots ,h_{\ell-1}\}$ be a subgroup of index 2 in $G.$ Additionally, let $\beta \notin H$ be an element in $G$ with $\beta^{-1}=\beta.$ List the elements of $G$ as in (\ref{fixed}), then any linear group matrix ring code in $(M_k(R))^n$ (a left ideal in $M_k(R)G$) is a reversible code.
\begin{proof}
Let $C_k(v)$ be a linear group matrix ring code in $(M_k(R))^n$ and use the ordering of the elements of $G$ given in (\ref{fixed}). If $(A_0,A_1,\dots ,A_{n-1}) \in C_k(v),$ then the following element is in $\Omega(C_k(v))$:
\begin{align*}
v= & A_0e+A_1h_1+A_2h_2+\dots+A_{\ell-1}h_{\ell-1}+A_{\ell}\beta h_{\ell-1}+A_{\ell-1}\beta h_{\ell-2}+\dots\\
+& A_{2\ell-3}\beta h_2+A_{2\ell-2} \beta h_1+A_{2\ell-1} \beta.
\end{align*}
Then $\beta v \in \Omega(C_k(v)).$ We have
\begin{align*}
\beta v= & A_0\beta e+A_1\beta h_1+A_2\beta h_2+\dots+A_{\ell-1}\beta h_{\ell-1}+A_{\ell}\beta \beta h_{\ell-1}+A_{\ell-1}\beta \beta h_{\ell-2}+\dots\\
+& A_{2\ell-3}\beta \beta h_2+A_{2\ell-2} \beta \beta h_1+A_{2\ell-1}\beta \beta\\
=& A_{2\ell-1}e+A_{2\ell-2}h_1+A_{2\ell-3}h_2+\dots +A_2\beta h_2+A_1\beta h_1+A_0e\beta.
\end{align*}
This gives that $(A_{n-1},A_{n-2},A_{n-3},\dots,A_2,A_1,A_0) \in C_k(v).$
\end{proof}
\end{theorem}

We now state the main result of this work.

\begin{theorem}\label{MainTheorem}
Let $R$ be a finite commutative Frobenius ring, $k$ be a positive integer and $G$ be a finite group of order $n=2\ell.$ Also, let $H=\{e,h_1,h_2,\dots ,h_{\ell-1}\}$ be a subgroup of index 2 in $G.$ Additionally, let $\beta \notin H$ be an element in $G$ with $\beta^{-1}=\beta.$ List the elements of $G$ as in (\ref{fixed}). Next, let $T_1, T_2,\dots, T_n$ be finite groups, each of order $k=2m$ and let $S_i=\{e, (s_i)_1, (s_i)_2, \dots, (s_i)_{m-1}\}$ be a subgroup of index $2$ in $T_i$.
Let $\mu \notin S_i$ be an element in $T_i$, with $\mu^{-1}=\mu.$ List the elements of $T_i$ as in (\ref{fixed2}). Then the code $B_k^*(v)$ in $R^{kn}$ is a reversible code of index 1.
\begin{proof}
From Theorem~\ref{RevGroupMatrixRingCode} we have that the code $C_k^*(v)$ (of length $n$) is reversible. From Theorem~\ref{RevGroupCode} we have that each block $\sigma(v_{g_i})$ (of length $k$) is reversible of index 1. Let $a_i^j \in R,$ where $i=\{1,2,\dots,k\}, j=\{1,2,\dots,n\},$ then it follows that given
$$((a_1^1,a_2^1,\dots,a_{k}^1),(a_1^2,a_2^2,\dots,a_{k}^2),\dots,(a_1^n,a_2^n,\dots,a_{k}^n)) \in B_k^*(v)$$
implies that
$$((a_{k}^n,a_{k-1}^n,\dots,a_{1}^n),(a_{k}^{n-1},a_{k-1}^{n-1},\dots,a_{1}^{n-1}),\dots,(a_{k}^1,a_{k-1}^1,\dots,a_{1}^1)) \in B_k^*(v).$$
Thus, the code $B_k^*(v)$ in $R^{kn}$ is a reversible code of index 1.
\end{proof}
\end{theorem}

\subsection{Applications to DNA Codes}

We now present an application of Theorem~\ref{MainTheorem}. Namely, we show that one can construct DNA codes with this theorem. Since Theorem~\ref{MainTheorem} guarantees reversibility, which is the basic property of a DNA code, we can combine this result with the map $\eta$ given in Section~\ref{LinearandDNACodes} to get the following.

\begin{theorem}\label{DNATheorem}
Let $R=\mathbb{F}_4,$ $k$ be a positive integer and $G$ be a finite group of order $n=2\ell.$ Also, let $H=\{e,h_1,h_2,\dots ,h_{\ell-1}\}$ be a subgroup of index 2 in $G.$ Additionally, let $\beta \notin H$ be an element in $G$ with $\beta^{-1}=\beta.$ List the elements of $G$ as in (\ref{fixed}). Next, let $T_1, T_2,\dots, T_n$ be finite groups, each of order $k=2m$ and let $S_i=\{e, (s_i)_1, (s_i)_2, \dots, (s_i)_{m-1}\}$ be a subgroup of index $2$ in $T_i$.
Let $\mu \notin S_i$ be an element in $T_i$, with $\mu^{-1}=\mu.$ List the elements of $T_i$ as in (\ref{fixed2}). Then the code $\eta (B_k^*(v))$ in $R^{kn}$ is a reversible DNA code of index 1.
\end{theorem}

We now give one possible matrix construction for $\tau_k^*(v)$ that we then employ to construct DNA codes with. We particularly employ the dihedral group with an even number of elements. We start with a definition.

\begin{definition}
Define the dihedral group as $D_{2p}=\langle a,b \ | \ a^p=b^2=e, \ aba=b^{-1}\rangle.$ We now list the elements of $D_{2p}$ according to Equation~(\ref{fixed}):
\begin{equation}\label{dihedralfixed}
\{e,a,a^2,\dots,a^{p-1},ba^{p-1},ba^{p-2},\dots,ba,b\}.
\end{equation}
\end{definition}
From now on, every-time we employ a dihedral group, we list its elements as in Equation~(\ref{dihedralfixed}). We now give one possible matrix construction for $\tau_k^*(v).$

Let 
$$G=D_{2\ell}=\{g_1,g_2,\dots,g_{n=2\ell}\}$$
and let 
$$T=D_{2m}=\{t_1,t_2,\dots,t_{k=2m}\}.$$ 
Next, let 
$$v=\sigma(v_{g_1})g_1+\sigma(v_{g_2})g_2+\dots+\sigma(v_{g_2\ell})g_{2\ell} \in M_{k}(R)G, $$
where
$$v_{g_i}=\alpha_{i_1}t_1+\alpha_{i_2}t_2+\dots+\alpha_{i_{2m}}t_{2m} \in RT \ \text{for} \ i=\{1,2,\dots,2\ell\}.$$
Then
$$\mathcal{G}=\tau_{k}^*(v)=$$
\begin{equation}
\begin{pmatrix}
\sigma(v_{g_1^{-1}g_1})&\sigma(v_{g_1^{-1}g_2})&\sigma(v_{g_1^{-1}g_3})&\dots&\sigma(v_{g_1^{-1}g_n})\\
\sigma(v_{g_1^{-2}g_1})&\sigma(v_{g_2^{-1}g_2})&\sigma(v_{g_2^{-1}g_3})&\dots&\sigma(v_{g_2^{-1}g_n})\\
\vdots&\vdots&\vdots&\vdots&\vdots\\
\sigma(v_{g_n^{-2}g_1})&\sigma(v_{g_n^{-1}g_2})&\sigma(v_{g_n^{-1}g_3})&\dots&\sigma(v_{g_n^{-1}g_n})\\
\end{pmatrix},
\end{equation}
where
$$\sigma(v_{g_i})=\begin{pmatrix}
A_i&B_i\\
B_i^T&A_i^T
\end{pmatrix}$$
with $A_i=circ(\alpha_{i_1},\alpha_{i_2},\dots,\alpha_{i_m}), B_i=circ(\alpha_{i_{m+1}},\alpha_{i_{m+2}},\dots,\alpha_{i_{2m}}).$

It can be easily observed that the matrix $\mathcal{G}$ is reversible, i.e., the reverse of each row of $\mathcal{G}$ is in $\mathcal{G}.$ We want to stress that this is only one possible construction for the matrix $\tau_{k}^*(v)$ that is reversible. There are many other groups that one could consider to define a matrix $\tau_{k}^*(v)$ with, that is reversible. An advantage of the matrix construction given above is that we can employ it to construct DNA codes of different lengths - this is because the matrix $\mathcal{G}$ depends on the cardinality of the group $G$ and the cardinality of the group $T.$  In the next section, we employ the matrix $\mathcal{G}$ to search for DNA codes of different lengths, that is, we consider different values for $n$ and $k.$

\section{Computational Results}

In this section, we employ Theorem~\ref{DNATheorem} and the matrix construction $\mathcal{G}$ from the previous section to search for DNA codes with. Namely, we form the code $\langle \mathcal{G}=\tau_k^*(v) \rangle$ by taking all linear combinations of the rows of the matrix $\mathcal{G}$ with coefficients in $\mathbb{F}_4.$ We then use the bijective correspondence between $\mathbb{F}_4$ and the DNA alphabet to obtain DNA codes. For each DNA code that we construct, we only tabulate the length, the fixed Hamming distance $d$ and the lower bound on the maximum size of the DNA code for a fixed length $n$ and the fixed Hamming distance $d.$ Many of our lower bounds are better than the currently known best bounds. The bounds that are equal to or better than the currently known best bounds are written in bold. Similarly, any new results are also written in bold. We perform our search in the software package MAGMA (\cite{MAGMA}) using a heuristic search scheme called the virus optimization algorithm (VOA). This method, as shown in \cite{Korban2}, allows one to obtain the computational results significantly faster then the standard linear search. With this in mind, by combining our construction method and the heuristic search scheme, we are able to obtain DNA codes with lengths higher than $42$ (this length is the highest known in the literature) that satisfy the Hamming distance, reverse, reverse-complement and the fixed GC-content constraints, in a very short period of time. For more details on this approach, please see \cite{Korban2}. The generator matrices, weight enumerators, GC-weight enumerators for the codes constructed can be found at \cite{GenMatr.}.

In Table~\ref{Table1}, we list the DNA codes obtained from $\langle \mathcal{G}=\tau_{k}^*(v) \rangle$ with $n=4$ and $k=4.$ 

\begin{table}[h!]
\caption{Lower bounds on $A_4^{RC}(n,d)$ and $A_4^{RC,GC}(n,d,w)$}
\resizebox{1\textwidth}{!}{\begin{minipage}{\textwidth}\label{Table1}
\centering
\begin{tabular}{cccccc}
\hline
$n$  & $d$  & $A_4^{RC}(n,d)$ & Best known & $A_4^{RC,GC}(n,d,w)$ & Best known \\ \hline
$16$ & $4$  & $65536$         & $8386560$ \cite{TulpanI}  & $33152$              & $3293600$ \cite{GaboritI} \\
$16$ & $5$   & $65536$         & $411840$ \cite{GaboritI}  & $26720$              & $55376$ \cite{GaboritI}    \\
$16$ & $6$   & $65536$         & $130560$ \cite{TulpanI}   & $26720$              & $55424$ \cite{Aboluion1}    \\
$16$ & $7$   & $4096$          & $32640$ \cite{TulpanI}    & $2496$               & $13856$ \cite{Aboluion1}   \\
$16$ & $8$   & $4096$          & $6680$ \cite{GaboritI}     & $1728$               & $3776$ \cite{Aboluion1}     \\
$16$ & $9$   & $256$           & $532$ \cite{GaboritI}     & $60$                 & $243$ \cite{Aboluion1}      \\
$16$ & $11$  & $\mathbf{256}$           & $120$ \cite{TulpanI}     & $60$        & $68$ \cite{Aboluion1}       \\ \hline
\end{tabular}
\end{minipage}}
\end{table}

In Table~\ref{Table2}, we list the DNA codes obtained from $\langle \mathcal{G}=\tau_{k}^*(v) \rangle$ with $n=4$ and $k=6.$ 
\begin{table}[h!]
\caption{Lower bounds on $A_4^{RC}(n,d)$ and $A_4^{RC,GC}(n,d,w)$}
\resizebox{1\textwidth}{!}{\begin{minipage}{\textwidth}\label{Table2}
\centering
\begin{tabular}{cccccc}
\hline
$n$  & $d$  & $A_4^{RC}(n,d)$ & Best known & $A_4^{RC,GC}(n,d,w)$ & Best known \\ \hline
$24$ & $4$  & $4294967296$         & $68719476736$ \cite{ReversibleDNACodes}   & $1400569856$              & $44319794176$ \cite{Aboluion1}  \\
$24$ & $5$  & $\mathbf{4294967296}$         & $-$   & $\mathbf{1385000960}$              & $346436544$ \cite{Aboluion1}  \\
$24$ & $6$  & $\mathbf{4294967296}$         & $268435456$ \cite{ReversibleDNACodes}   & $\mathbf{1385947136}$              & $43355616$ \cite{Aboluion1} \\
$24$ & $8$  & $\mathbf{16777216}$         & $8386560$ \cite{Kim1}   & $\mathbf{5712896}$              & $5406464$ \cite{Aboluion1} \\
 \hline
\end{tabular}
\end{minipage}}
\end{table}

In Table~\ref{Table3}, we list the DNA codes obtained from $\langle \mathcal{G}=\tau_{k}^*(v) \rangle$ with $n=4$ and $k=8.$ 
\begin{table}[h!]
\caption{Lower bounds on $A_4^{RC}(n,d)$ and $A_4^{RC,GC}(n,d,w)$}
\resizebox{0.9\textwidth}{!}{\begin{minipage}{\textwidth}\label{Table3}
\centering
\begin{tabular}{cccccc}
\hline
$n$  & $d$ & $A_4^{RC}(n,d)$ & Best known & $A_4^{RC,GC}(n,d,w)$ & Best known \\ \hline
$32$ & $4$ & $1099511627776$         & $17592186044416$ \cite{ReversibleDNACodes}     & $308354940928$              & $4928618364928$  \cite{ReversibleDNACodes} \\
$32$ & $6$ & $\mathbf{68719476736}$         & $-$   & $\mathbf{19321872384}$              & $-$  \\
$32$ & $8$ & $\mathbf{4294967296}$         & $-$   & $\mathbf{1207617024}$              & $-$  \\
 \hline
\end{tabular}
\end{minipage}}
\end{table}
\newpage
In Table~\ref{Table4}, we list the DNA codes obtained from $\langle \mathcal{G}=\tau_{k}^*(v) \rangle$ with $n=6$ and $k=6.$
\begin{table}[h!]

\caption{Lower bounds on $A_4^{RC}(n,d)$ and $A_4^{RC,GC}(n,d,w)$}
\resizebox{0.845\textwidth}{!}{\begin{minipage}{\textwidth}\label{Table4}
\centering
\begin{tabular}{cccccc}
\hline
$n$  & $d$ & $A_4^{RC}(n,d)$ & Best known & $A_4^{RC,GC}(n,d,w)$ & Best known \\ \hline
$36$ & $4$ & $\mathbf{1152921504606846976}$         & $-$   & $\mathbf{304509378898165760}$              & $-$  \\
$36$ & $5$ & $\mathbf{4503599627370496}$         & $-$   & $\mathbf{1189489761320960}$              & $-$  \\
$36$ & $6$ & $\mathbf{4503599627370496}$         & $-$   & $\mathbf{1189496134041600}$              & $-$  \\
 \hline
\end{tabular}
\end{minipage}}
\end{table}

In Table~\ref{Table5}, we list the DNA codes obtained from $\langle \mathcal{G}=\tau_{k}^*(v) \rangle$ with $n=4$ and $k=12.$
\begin{table}[h!]
\caption{Lower bounds on $A_4^{RC}(n,d)$ and $A_4^{RC,GC}(n,d,w)$}
\resizebox{0.75\textwidth}{!}{\begin{minipage}{\textwidth}\label{Table5}
\centering
\begin{tabular}{cccccc}
\hline
$n$  & $d$ & $A_4^{RC}(n,d)$ & Best known & $A_4^{RC,GC}(n,d,w)$ & Best known \\ \hline
$48$ & $4$ & $\mathbf{75557863725914323419136}$         & $-$   & $\mathbf{17312857989983938543616}$              & $-$  \\
$48$ & $6$ & $\mathbf{295147905179352825856}$         & $-$   & $\mathbf{67628166256403677184}$              & $-$  \\
$48$ & $8$ & $\mathbf{18446744073709551616}$         & $-$   & $\mathbf{4226771970210922496}$              & $-$  \\
 \hline
\end{tabular}
\end{minipage}}
\end{table}

In Table~\ref{Table6}, we list the DNA codes obtained from $\langle \mathcal{G}=\tau_{k}^*(v) \rangle$ with $n=4$ and $k=14.$ 
\begin{table}[h!]
\caption{Lower bounds on $A_4^{RC}(n,d)$ and $A_4^{RC,GC}(n,d,w)$}
\resizebox{0.6\textwidth}{!}{\begin{minipage}{\textwidth}\label{Table6}
\centering
\begin{tabular}{cccccc}
\hline
$n$  & $d$ & $A_4^{RC}(n,d)$ & Best known & $A_4^{RC,GC}(n,d,w)$ & Best known \\ \hline
$56$ & $2$ & $\mathbf{20282409603651670423947251286016}$         & $-$   & $\mathbf{4305830175517583306596737351680}$              & $-$  \\
$56$ & $4$ & $\mathbf{309485009821345068724781056}$         & $-$   & $\mathbf{32850875987488682530570240}$              & $-$  \\
$56$ & $7$ & $\mathbf{1208925819614629174706176}$         & $-$   & $\mathbf{256647469998345059041280}$              & $-$  \\
$56$ & $8$ & $\mathbf{1208925819614629174706176}$         & $-$   & $\mathbf{256647469998345059041280}$              & $-$  \\
 \hline
\end{tabular}
\end{minipage}}
\end{table}

In Table~\ref{Table7}, we list the DNA codes obtained from $\langle \mathcal{G}=\tau_{k}^*(v) \rangle$ with $n=6$ and $k=10.$ 
\newpage
\begin{table}[h!]
\caption{Lower bounds on $A_4^{RC}(n,d)$ and $A_4^{RC,GC}(n,d,w)$}
\resizebox{0.65\textwidth}{!}{\begin{minipage}{\textwidth}\label{Table7}
\centering
\begin{tabular}{cccccc}
\hline
$n$  & $d$ & $A_4^{RC}(n,d)$ & Best known & $A_4^{RC,GC}(n,d,w)$ & Best known \\ \hline
$60$ & $6$ & $\mathbf{79228162514264337593543950336}$         & $-$   & $\mathbf{8127080161539269101631832064}$              & $-$  \\

 \hline
\end{tabular}
\end{minipage}}
\end{table}

In Table~\ref{Table8}, we list the DNA codes obtained from $\langle \mathcal{G}=\tau_{k}^*(v) \rangle$ with $n=4$ and $k=16.$ 
\begin{table}[h!]
\caption{Lower bounds on $A_4^{RC}(n,d)$ and $A_4^{RC,GC}(n,d,w)$}
\resizebox{0.675\textwidth}{!}{\begin{minipage}{\textwidth}\label{Table8}
\centering
\begin{tabular}{cccccc}
\hline
$n$  & $d$ & $A_4^{RC}(n,d)$ & Best known & $A_4^{RC,GC}(n,d,w)$ & Best known \\ \hline
$64$ & $8$ & $\mathbf{309485009821345068724781056}$         & $-$   & $\mathbf{61492671022720836698112000}$              & $-$  \\

 \hline
\end{tabular}
\end{minipage}}
\end{table}

In Table~\ref{Table9}, we list the DNA codes obtained from $\langle \mathcal{G}=\tau_{k}^*(v) \rangle$ with $n=4$ and $k=18.$ 
\begin{table}[h!]
\caption{Lower bounds on $A_4^{RC}(n,d)$ and $A_4^{RC,GC}(n,d,w)$}
\resizebox{0.585\textwidth}{!}{\begin{minipage}{\textwidth}\label{Table9}
\centering
\begin{tabular}{cccccc}
\hline
$n$  & $d$ & $A_4^{RC}(n,d)$ & Best known & $A_4^{RC,GC}(n,d,w)$ & Best known \\ \hline
$72$ & $4$ & $\mathbf{5192296858534827628530496329220096}$         & $-$   & $\mathbf{973095367002024438257652919697408}$              & $-$  \\
$72$ & $6$ & $\mathbf{5192296858534827628530496329220096}$         & $-$   & $\mathbf{973095366962111571113691167326208}$              & $-$  \\
 \hline
\end{tabular}
\end{minipage}}
\end{table}

We note that the lower bounds on $A_4^{RC}(n,d)$ and $A_4^{RC,GC}(n,d,w)$ for $n=48, 56, 60, 64, 72$ have not been known in the literature before. Only lower bounds for codes of length 42 or less have been considered. Additionally, the lower bounds for some codes of lengths less than 42, for fixed parameters $n$ and $d,$ are also new to the literature. 

We now present an example of how we construct the DNA codes using our group matrix ring approach.

\begin{example}
Let $D_4$ be the dihedral group of order 4 with the ordering of its elements given as $\{e, a,  ba, b \}.$ Consider the group matrix ring element 
$$v= \sigma(v_1)e+ \sigma(v_2)a+ \sigma(v_1)ba+ \sigma(v_1)b \in M_2(\mathbb{F}_4)D_4,$$  
where,
$$\sigma(v_1)=circ(0, w^2), ~~\sigma(v_2)=circ(w, w^2), ~~\sigma(v_3)=circ(w, 1), ~~\sigma(v_4)=circ(0, 1).$$

Then the generator matrix has the following form:

\begin{equation}\label{examplegc}
\mathcal{G}=\tau_k^*(v)=
\begin{pmatrix}
0& w^2 & w & w^2& w & 1 &0 & 1 \\
w^2 &  0& w^2&   w&   1&   w &  1&   0 \\
 w &w^2   &0 &w^2   &0   &1 &  w   &1 \\
w^2 &  w& w^2&   0&   1&   0&   1&   w \\
 w  & 1  & 0  & 1  & 0 &w^2   &w & w^2 \\
1 &  w &  1 &  0 &w^2 &  0 & w^2 &  w \\
0  & 1  & w  & 1  & w & w^2  & 0 & w^2\\
1  & 0 &  1 &  w & w^2 &  w &w^2 &  0\\
\end{pmatrix}
.
\end{equation}

Now, using the generator matrix $\mathcal{G}$, we construct a DNA code $D$ with 256 codewords that satisfy the Hamming distance and the $RC$-constraints with $d=4.$ The codewords of this code are given in Table \ref{exrc}.

\begin{table}[htbp]
\caption{A DNA code of length 8 with 256 codewords satisfying the HD and RC constraints}
\resizebox{0.80\textwidth}{!}{\begin{minipage}{\textwidth}\label{exrc}
\centering
\begin{tabular}{ccccccc}
\hline
AAAAAAAA &TAAAATTT &CAAAACCC &GAAAAGGG &GTAATGCC &CTAATCGG\\
TTAATTAA &ATAATATT &ACAACACC &TCAACTGG &CCAACCAA &GCAACGTT\\
GGAAGGAA &CGAAGCTT &TGAAGTCC &AGAAGAGG &AGTACTGC &TGTACACG\\
CGTACGTA &GGTACCAT &GCTAGCTA &CCTAGGAT &TCTAGAGC &ACTAGTCG\\
ATTAATTA &TTTAAAAT &CTTAAGGC &GTTAACCG &GATATCGC &CATATGCG\\
TATATATA &AATATTAT &AACACCAC &TACACGTG &CACACACA &GACACTGT\\
GTCAGTCA &CTCAGAGT &TTCAGGAC &ATCAGCTG &ACCAACCA &TCCAAGGT\\
CCCAAAAC &GCCAATTG &GGCATTAC &CGCATATG &TGCATGCA &AGCATCGT\\
AGGAAGGA &TGGAACCT &CGGAATTC &GGGAAAAG &GCGATATC &CCGATTAG\\
TCGATCGA &ACGATGCT &ATGACGTC &TTGACCAG &CTGACTGA &GTGACACT\\
GAGAGAGA &CAGAGTCT &TAGAGCTC &AAGAGGAG &AAGTCCTG &TAGTCGAC\\
CAGTCAGT &GAGTCTCA &GTGTGTGT &CTGTGACA &TTGTGGTG &ATGTGCAC\\
ACGTACGT &TCGTAGCA &CCGTAATG &GCGTATAC &GGGTTTTG &CGGTTAAC\\
TGGTTGGT &AGGTTCCA &AGCTAGCT &TGCTACGA &CGCTATAG &GGCTAATC\\
GCCTTAAG &CCCTTTTC &TCCTTCCT &ACCTTGGA &ATCTCGAG &TTCTCCTC\\
CTCTCTCT &GTCTCAGA &GACTGACT &CACTGTGA &TACTGCAG &AACTGGTC\\
AATTAATT &TATTATAA &CATTACGG &GATTAGCC &GTTTTGGG &CTTTTCCC\\
TTTTTTTT &ATTTTAAA &ACTTCAGG &TCTTCTCC &CCTTCCTT &GCTTCGAA\\
GGTTGGTT &CGTTGCAA &TGTTGTGG &AGTTGACC &AGATCTCG &TGATCAGC\\
CGATCGAT &GGATCCTA &GCATGCAT &CCATGGTA &TCATGACG &ACATGTGC\\
ATATATAT &TTATAATA &CTATAGCG &GTATACGC &GAATTCCG &CAATTGGC\\
TAATTAAT &AAATTTTA &AAACCCCA &TAACCGGT &CAACCAAC &GAACCTTG\\
GTACGTAC &CTACGATG &TTACGGCA &ATACGCGT &ACACACAC &TCACAGTG\\
CCACAACA &GCACATGT &GGACTTCA &CGACTAGT &TGACTGAC &AGACTCTG\\
AGTCAGTC &TGTCACAG &CGTCATGA &GGTCAACT &GCTCTAGA &CCTCTTCT\\
TCTCTCTC &ACTCTGAG &ATTCCGGA &TTTCCCCT &CTTCCTTC &GTTCCAAG\\
GATCGATC &CATCGTAG &TATCGCGA &AATCGGCT &AACCAACC &TACCATGG\\
CACCACAA &GACCAGTT &GTCCTGAA &CTCCTCTT &TTCCTTCC &ATCCTAGG\\
ACCCCAAA &TCCCCTTT &CCCCCCCC &GCCCCGGG &GGCCGGCC &CGCCGCGG\\
TGCCGTAA &AGCCGATT &AGGCCTTA &TGGCCAAT &CGGCCGGC &GGGCCCCG\\
GCGCGCGC &CCGCGGCG &TCGCGATA &ACGCGTAT &ATGCATGC &TTGCAACG\\
CTGCAGTA &GTGCACAT &GAGCTCTA &CAGCTGAT &TAGCTAGC &AAGCTTCG\\
AAGGAAGG &TAGGATCC &CAGGACTT &GAGGAGAA &GTGGTGTT &CTGGTCAA\\
TTGGTTGG &ATGGTACC &ACGGCATT &TCGGCTAA &CCGGCCGG &GCGGCGCC\\
GGGGGGGG &CGGGGCCC &TGGGGTTT &AGGGGAAA &AGCGCTAT &TGCGCATA\\
CGCGCGCG &GGCGCCGC &GCCGGCCG &CCCGGGGC &TCCGGAAT &ACCGGTTA\\
ATCGATCG &TTCGAAGC &CTCGAGAT &GTCGACTA &GACGTCAT &CACGTGTA\\
TACGTACG &AACGTTGC &AATGCCGT &TATGCGCA &CATGCATG &GATGCTAC\\
GTTGGTTG &CTTGGAAC &TTTGGGGT &ATTGGCCA &ACTGACTG &TCTGAGAC\\
CCTGAAGT &GCTGATCA &GGTGTTGT &CGTGTACA &TGTGTGTG &AGTGTCAC\\
AGAGAGAG &TGAGACTC &CGAGATCT &GGAGAAGA &GCAGTACT &CCAGTTGA\\
TCAGTCAG &ACAGTGTC &ATAGCGCT &TTAGCCGA &CTAGCTAG &GTAGCATC\\
GAAGGAAG &CAAGGTTC &TAAGGCCT &AAAGGGGA & &\\ \\ \hline
\end{tabular}
\end{minipage}}
\end{table}
\newpage
The GC-weight enumerator of $D$ is
$$GCW(a,b)=16a^8 + 224a^4b^4 + 16b^8.$$
Thus, we can now construct a DNA code that satisfies the Hamming distance constraint with $d=4,$ the reversible complement constraint and the
fixed GC-content constraint with $w=4.$ Such DNA code has 224 codewords. The elements of this DNA code are given in Table \ref{exgc}.

\begin{table}[htbp]
\caption{A DNA code of length 8 with 224 codewords satisfying the HD, RC and GC constraints}
\resizebox{0.80\textwidth}{!}{\begin{minipage}{\textwidth}\label{exgc}
\centering
\begin{tabular}{ccccccc}
\hline
TAAGGCCT &AAAGGGGA &CAAAACCC &GAAAAGGG &GTAATGCC &CTAATCGG\\
GAAGGAAG &CAAGGTTC &ACAACACC &TCAACTGG &CCAACCAA &GCAACGTT\\
GGAAGGAA &CGAAGCTT &TGAAGTCC &AGAAGAGG &AGTACTGC &TGTACACG\\
CGTACGTA &GGTACCAT &GCTAGCTA &CCTAGGAT &TCTAGAGC &ACTAGTCG\\
TCCGGAAT &ACCGGTTA &CTTAAGGC &GTTAACCG &GATATCGC &CATATGCG\\
ACCCCAAA &TCCCCTTT &AACACCAC &TACACGTG &CACACACA &GACACTGT\\
GTCAGTCA &CTCAGAGT &TTCAGGAC &ATCAGCTG &ACCAACCA &TCCAAGGT\\
CCCAAAAC &GCCAATTG &GGCATTAC &CGCATATG &TGCATGCA &AGCATCGT\\
AGGAAGGA &TGGAACCT &CGGAATTC &GGGAAAAG &GCGATATC &CCGATTAG\\
TCGATCGA &ACGATGCT &ATGACGTC &TTGACCAG &CTGACTGA &GTGACACT\\
GAGAGAGA &CAGAGTCT &TAGAGCTC &AAGAGGAG &AAGTCCTG &TAGTCGAC\\
CAGTCAGT &GAGTCTCA &GTGTGTGT &CTGTGACA &TTGTGGTG &ATGTGCAC\\
ACGTACGT &TCGTAGCA &CCGTAATG &GCGTATAC &GGGTTTTG &CGGTTAAC\\
TGGTTGGT &AGGTTCCA &AGCTAGCT &TGCTACGA &CGCTATAG &GGCTAATC\\
GCCTTAAG &CCCTTTTC &TCCTTCCT &ACCTTGGA &ATCTCGAG &TTCTCCTC\\
CTCTCTCT &GTCTCAGA &GACTGACT &CACTGTGA &TACTGCAG &AACTGGTC\\
TGGGGTTT &AGGGGAAA &CATTACGG &GATTAGCC &GTTTTGGG &CTTTTCCC\\
AGCGCTAT &TGCGCATA &ACTTCAGG &TCTTCTCC &CCTTCCTT &GCTTCGAA\\
GGTTGGTT &CGTTGCAA &TGTTGTGG &AGTTGACC &AGATCTCG &TGATCAGC\\
CGATCGAT &GGATCCTA &GCATGCAT &CCATGGTA &TCATGACG &ACATGTGC\\
TTGGTTGG &ATGGTACC &CTATAGCG &GTATACGC &GAATTCCG &CAATTGGC\\
ACGGCATT &TCGGCTAA&AAACCCCA &TAACCGGT &CAACCAAC &GAACCTTG\\
GTACGTAC &CTACGATG &TTACGGCA &ATACGCGT &ACACACAC &TCACAGTG\\
CCACAACA &GCACATGT &GGACTTCA &CGACTAGT &TGACTGAC &AGACTCTG\\
AGTCAGTC &TGTCACAG &CGTCATGA &GGTCAACT &GCTCTAGA &CCTCTTCT\\
TCTCTCTC &ACTCTGAG &ATTCCGGA &TTTCCCCT &CTTCCTTC &GTTCCAAG\\
GATCGATC &CATCGTAG &TATCGCGA &AATCGGCT &AACCAACC &TACCATGG\\
CACCACAA &GACCAGTT &GTCCTGAA &CTCCTCTT &TTCCTTCC &ATCCTAGG\\
TGCCGTAA &AGCCGATT &AGGCCTTA &TGGCCAAT &TCGCGATA &ACGCGTAT\\
CTGCAGTA &GTGCACAT &GAGCTCTA &CAGCTGAT &TAGCTAGC &AAGCTTCG\\
AAGGAAGG &TAGGATCC &CAGGACTT &GAGGAGAA &GTGGTGTT &CTGGTCAA\\
ATCGATCG &TTCGAAGC &CTCGAGAT &GTCGACTA &GACGTCAT &CACGTGTA\\
TACGTACG &AACGTTGC &AATGCCGT &TATGCGCA &CATGCATG &GATGCTAC\\
GTTGGTTG &CTTGGAAC &TTTGGGGT &ATTGGCCA &ACTGACTG &TCTGAGAC\\
CCTGAAGT &GCTGATCA &GGTGTTGT &CGTGTACA &TGTGTGTG &AGTGTCAC\\
AGAGAGAG &TGAGACTC &CGAGATCT &GGAGAAGA &GCAGTACT &CCAGTTGA\\
TCAGTCAG &ACAGTGTC &ATAGCGCT &TTAGCCGA &CTAGCTAG &GTAGCATC\\
ATGCATGC &TTGCAACG &          &         &        &         \\ \\ \hline
\end{tabular}
\end{minipage}}
\end{table}

\end{example}

\newpage
\section{Conclusion}

In this paper, we presented a method for constructing DNA codes. Our method uses group matrix rings and a known in the literature matrix construction. We showed that with our new construction method, one can obtain reversible codes of length $kn,$ where $k, n \in \mathbb{N},$ over the finite commutative Frobenius ring $R.$ We constructed many DNA codes that satisfy the Hamming distance, reverse, reverse-complement and the fixed GC-content constraints. Moreover, we improved many lower bounds on the sizes of some known DNA codes and we also gave new lower bounds on the sizes of some DNA codes of lengths $48, 56, 60, 64$ and $72$ for some fixed values of the Hamming distance $d.$ A possible direction for future research is to consider our approach for groups different than the dihedral group. We believe that by considering other groups of different lengths, one can obtain more interesting DNA codes that satisfy the above mentioned constraints.


\begin{thebibliography}{99}

\bibitem{Aboluion1} N. Aboluion, D. H. Smith, S. Perkins, \lq \lq Linear and nonlinear constructions of DNA codes with Hamming distance d, constant GC-content and a reverse-complement constraint\rq \rq, Discrete Math., vol. 312, pp. 1062--1075, 2012.

\bibitem{Abualrub1} T. Abualrub, A. Ghrayeb, X.N. Zeng, \lq \lq Construction of cyclic codes over $GF(4)$ for DNA computing \rq \rq, Journal of the Franklin Institute, vol. 343, pp. 448--457, 2006.

\bibitem{Adleman} L. Adleman, \lq \lq Molecular computation of the solutions to combinatorial problems", Science, vol. 266, pp. 1021--1024, 1994.

\bibitem{Adleman2} L. Adleman, P.W.K. Rothemund, S. Rowies, E. Winfree, \lq \lq On applying molecular computation to the data encryption standard", J. Comp. Biology, vol. 6, pp. 53--63, 1999.

\bibitem{Boneh1} D. Boneh, C. Dunworth and R. Lipton, \lq \lq Breaking DES using molecular computer", Princenton CS Tech-Report, Number CS--TR--489--95, 1995.

\bibitem{MAGMA} W. Bosma, J. Cannon and C. Playoust, \lq \lq The Magma algebra system. I. The user language", J. Symbolic Comput., vol. 24, pp. 235--265, 1997.

\bibitem{Blawat1} M. Blawat, et al., \lq \lq Forward error correction for DNA data storage\rq \rq, Procedia Comput. Sci., vol. 80, pp. 1011--1022, 2016.

\bibitem{ReversibleDNACodes} Y. Cengellenmis, A. Dertli, S.T. Dougherty, A. Korban, S. Sahinkaya, D. Ustun, \lq \lq Reversible $G$-Codes over the Ring $\mathcal{F}_{j,k}$ with Applications to DNA Codes", \textbf{in submission}.

\bibitem{Church1} G. M. Church, Y. Gao, S. Kossuri, \lq \lq Next-generation digital information storage in DNA\rq \rq, Science, vol 337, pp. 1628, 2012.

\bibitem{Dougherty1} S.T. Dougherty, J. Gildea, R. Taylor and A.  Tylshchak, \lq \lq Group rings, $G$-codes and constructions of self-dual and formally self-dual codes", Designs, Codes and Cryptography, vol. 86, pp. 2115--2138, 2018.

\bibitem{M(R)GPaper} S.T. Dougherty, A. Korban, S. Sahinkaya, D. Ustun, \lq \lq Group Matrix Ring Codes and Constructions of Self-Dual Codes \rq \rq, Applicable Algebra in Engineering, Communication and Computing, doi:https://doi.org/10.1007/s00200-021-00504-9.

\bibitem{GaboritI} P. Gaborit, O. D. King, \lq \lq Linear Constructions for DNA Codes\rq \rq, Theoretical Computer Science, vol. 334, pp. 99--113, 2005.

\bibitem{HurleyI} T. Hurley, \lq \lq Group Rings and Rings of Matrices", Int. Jour. Pure
and Appl. Math, vol. 31, no. 3, pp. 319--335, 2006.

\bibitem{Kim1} H. J. Kim, W-H. Choi, Y. Lee, \lq \lq Designing DNA codes from reversible self-dual codes over $GF(4)$\rq \rq, Discrete Mathematics, vol. 344, 2021.

\bibitem{King1} O.D. King, \lq \lq Bounds for DNA codes with constant GC-content\rq \rq, Electron. J. Comb., vol. 10, pp. 33, 2003.

\bibitem{GenMatr.} A. Korban, S. Sahinkaya, D. Ustun, Generator Matrices for the manuscript entitled \lq \lq Reversible $G^k$-Codes with Applications to DNA Codes", available at https://sites.google.com/view/adriankorban/generator-matrices.

\bibitem{Korban2} A. Korban, S. Sahinkaya, D. Ustun, \lq \lq An Application of the Virus Optimization Algorithm to the Problem of Finding Extremal Binary Self-Dual Codes ", arXiv:2103.07739v1.

\bibitem{Limbachiya1} D. Limbachiya, B. Rao, M. K. Gupta, \lq \lq The art of DNA strings\rq \rq, Sixteen years of DNA coding theory, CoRR, vol. abs/1607.00266 [Online], http://arxiv.org/abs/1607.00266, 2016.

\bibitem{Marathe1} A. Marathe, A.E. Condon, R.M. Corn, On combinatorial DNA word design, J. Comput. Biol., vol. 8, pp. 201--220, 2001.

\bibitem{Oztas1} E. S. Oztas, B. Yildiz, I. Siap, \lq \lq A novel approach for constructing reversible codes and applications to DNA codes over the ring $\mathbb{F}_2[u]/(u^{2k}-1)$\rq \rq, Finite Fields and Their Applications, vol. 46, pp. 217--234, 2017.

\bibitem{TulpanI} D. Tulpan, D. H. Smith, R. Montemanni, \lq \lq Thermodynamic post-processing versus GC-content pre-processing for DNA codes satisfying the Hamming distance and reverse-complement constraints\rq \rq, IEEE/ACM Trans. Comput. Biol. Bioinform., vol 11, pp. 441--452, 2014.


\end{thebibliography}
\end{document}